# Comment on "Solar Wind and Heavy Ion Properties of Interplanetary Coronal Mass Ejections" by M.J.Owens

Yu.I.Yermolaev, I.G.Lodkina

Space Research Institute, Russian Academy of Sciences, Moscow, Russia

A recent paper by Owens (2018) presents a statistical analysis of the properties of interplanetary coronal mass ejections (ICMEs) and variations in their compositions and ion charge-state signatures using data from the Advanced Composition Explorer (ACE) spacecraft. However, the article contains several serious flaws, which we will discuss here.

Many parts of the work (data sources, processing methodology and most of the results) are similar to those which were presented in corresponding parts of an article by Lynch et al. (2003). However, the article by Owens (2018) did not reference the article by Lynch et al. (2003). As a result, the article did not discuss newly uncovered data in comparison to that which was previously reported. Further, part of the article by Owens (2018) appears to replicate material presented in the article by Lynch et al. (2003).

The results obtained in work by Owens (2018) either partially or completely contradict those previously obtained regarding the mass and charge composition of plasma in magnetic clouds (see, for example, the generally accepted properties of magnetic clouds reviewed by Zurbuchen and Richardson [2006]). In particular, the following criteria were not met: (1) enhanced α/proton ratio $He^{2+}/H^{+} > 8\%$ (Hirshberg et al., 1972; Borrini et al., 1982), (2) elevated oxygen charge states $O^{7+}/O^{6+} > 1$ (Henke et al., 2001; Zurbuchen et al. 2003), and (3) unusually high Fe charge states $<Q>_{Fe} > 12$ (Bame et al., 1979; Lepri et al., 2001; Lepri and Zurbuchen, 2004). The article by Owens (2018) ignores these inconsistencies, which indicates the incorrectness of the event selection, data analysis and conclusions presented.

Although the work presents results that were derived from the processing of experimental data, the work lacks a description of the methodology used to obtain data and its accuracy. The work uses the term "1-sigma variations", but does not explain what the value means or how it was calculated. In our opinion, differences between values shown in Fig. 5 are significantly less than standard deviations given for corresponding parameters. Therefore, many statements regarding differences between parameters are statistically unreliable.

Thus, our assessment of data presented in Owens (2018) indicates that the work does not meet basic requirements of scientific ethics, and conclusions presented by its author have the potential to mislead readers.

## References

Bame, S.J., Asbridge, J.R., Feldman, W.C. et al. (1979). Solar wind heavy ions from flare-heated coronal plasma. *Sol Phys* **62,** 179–201 https://doi.org/10.1007/BF00150143

Borrini, G., Gosling, J. T., Bame, S. J., and Feldman, W. C.(1982), Helium abundance enhancements in the solar wind, *J. Geophys. Res.* **87**, 7370 https://doi.org/10.1029/JA087iA09p07370

Henke, T., J. Woch, R. Schwenn, U. Mall, G. Gloeckler, R. von Steiger, R. J. Forsyth, A. Balogh (2001) Ionization state and magnetic topology of coronal mass ejections, *J. Geophys. Res.* Volume 106, Issue A6, Pages 10597-10613  https://doi.org/10.1029/2000JA900176

Hirshberg, J., Bame, S. J., and Robbins, D. E.: (1972), Solar flares and solar wind helium enrichments: July 1965–July 1967, *Sol. Phys.* **23**, 467

Lepri, S. T., Zurbuchen, T. H., Fisk, L. A., Richardson, I. G., Cane, H. V., and Gloeckler, G.: 2001, Iron charge distribution as an identifier of interplanetary coronal mass ejections, *J. Geophys. Res.* **106**, 29,231 https://doi.org/10.1029/2001JA000014

Lepri, S. T., and Zurbuchen, T. H.: 2004, Iron charge state distributions as an indicator of hot ICMEs: Possible sources and temporal and spatial variations during solar maximum, *J. Geophys. Res.* **109**, doi:10.1029/2003JA009954.

Lynch, B. J., T. H. Zurbuchen, L. A. Fisk, and S. K. Antiochos (2003) Internal structure of magnetic clouds: Plasma and composition, J. Geophys. Res., 108(A6), 1239, doi:10.1029/2002JA009591

Owens, M. J. (2018) Solar Wind and Heavy Ion Properties of Interplanetary Coronal Mass Ejections, Solar Phys (2018) 293:122 https://doi.org/10.1007/s11207-018-1343-0

Zurbuchen, T., Fisk, L. A., Lepri, S. T., and von Steiger, R.: 2003, in Velli, M., Bruno, R., Malara, F. (eds.), *Solar Wind Ten*, AIP Conf. Proc. 679, Mellville, N. Y., p. 604.

Zurbuchen, T. H.; Richardson, I. G. (2006) In-situ solar wind and magnetic field signatures of interplanetary coronal mass ejections, Space Science Reviews (2006) 123: 31–43 DOI: 10.1007/s11214-006-9010-4